# COVIDX: Computer-aided diagnosis of Covid-19 and its severity prediction with raw digital chest X-ray images

Wajid Arshad Abbasi[1, *], Syed Ali Abbas[1], Saiqa Andleeb[2]

[1]Computaional Biology and Data Analysis Lab., Department of Computer Sciences & Information Technology, King Abdullah Campus, University of Azad Jammu & Kashmir, Muzaffarabad, AJ&K, 13100, Pakistan.

[2]Biotechnology Lab., Department of Zoology, King Abdullah Campus, University of Azad Jammu & Kashmir, Muzaffarabad, AJ&K, 13100, Pakistan.

*Corresponding author's emails: wajidarshad@gmail.com; wajidarshad@ajku.edu.pk

## Abstract

Coronavirus disease (COVID-19) is a contagious infection caused by severe acute respiratory syndrome coronavirus-2 (SARS-COV-2) and it has infected and killed millions of people across the globe. In the absence of specific drugs or vaccines for the treatment of COVID-19 and the limitation of prevailing diagnostic techniques, there is a requirement for some alternate automatic screening systems that can be used by the physicians to quickly identify and isolate the infected patients. A chest X-ray (CXR) image can be used as an alternative modality to detect and diagnose the COVID-19. In this study, we present an automatic COVID-19 diagnostic and severity prediction (COVIDX) system that uses deep feature maps from CXR images to diagnose COVID-19 and its severity prediction. The proposed system uses a three-phase classification approach (healthy vs unhealthy, COVID-19 vs Pneumonia, and COVID-19 severity) using different shallow supervised classification algorithms. We evaluated COVIDX not only through 10-fold cross-validation and by using an external validation dataset but also in real settings by involving an experienced radiologist. In all the evaluation settings, COVIDX outperforms all the existing state-of-the-art methods designed for this purpose. We made COVIDX easily accessible through a cloud-based webserver and python code available at https://sites.google.com/view/wajidarshad/software and https://github.com/wajidarshad/covidx, respectively.

## 1. Background

Coronavirus disease (COVID-19) is a contagious infection caused by severe acute respiratory syndrome coronavirus-2 (SARS-COV-2) and its transmission is also possible from asymptotic patients while incubation (Huang et al., 2020; Kooraki et al., 2020). This pandemic has infected and killed millions of people across the globe ("COVID-19 Map," n.d.). The World Health Organization (WHO) has already declared this pandemic as a global health calamity ("Coronavirus disease (COVID-19) – World Health Organization," n.d.). According to medical experts, this disease mainly infects the human respiratory system causing severe pneumonia showing symptoms of dry cough, breathing problems, fever, fatigue, and lung failure, etc (Cheng et al., 2020; Huang et al., 2020). Right now, the world is curiously waiting for a specific vaccine or medication to prevent this lethal infection.

In the absence of specific drugs or vaccines for the treatment of COVID-19, early diagnosis of the disease is crucial to avoid further spread by advising quarantine or isolation. Currently, there are two types of tests to detect COVID-19: diagnostic tests and antibody tests (Commissioner, 2020).

A diagnostic test such as Reverse transcription-polymerase chain reaction (RT-PCR) can be used to detect an active coronavirus infection. Whereas, the antibody test looks for antibodies created in your body in the response to the disease. RT-PCR is the most common diagnostic practice to detect viral infection (Sheikhzadeh et al., 2020). However, the standard confirmatory clinical RT-PCR test to detect COVID-19 is manual, complex, laborious, and costly (Chowdhury et al., 2020). Moreover, the limited availability of test kits and domain experts further hamper the situation. Meanwhile, antibody tests cannot be used to diagnose COVID-19 as after infection, antibodies can take numerous days or weeks to develop and may stay in your blood a bit longer after recovery (Commissioner, 2020). Keeping in view the limitation of prevailing diagnostic techniques and a rapid surge of infected patients, there is a requirement for some alternate automatic screening systems that can be used by the physicians to quickly identify and isolate the infected patients.

X-ray imaging is the most common noninvasive diagnostic technique that helps medical practitioners to diagnose and treat many diseases. A chest X-ray (CXR) is normally taken to assess the medical fitness of the lungs, heart, and chest wall. The CXR has also been playing a critical role in the pilot investigation of various respiratory irregularities (Chandra et al., 2021; Chandra and Verma, 2020a). In this context, a CXR image can also be used as an alternative modality to detect and diagnose the COVID-19. CXR images are normally interpreted by expert radiologists. Whereas, some studies show that CXR images of COVID-19 infected patients show diverse features (Cheng et al., 2020; Chowdhury et al., 2020; Zhang et al., 2020). Therefore, the manual interpretation of these CXR images with subtle variations is quite challenging. Moreover, the current enormous upsurge of infected patients makes it a challenging task for the domain experts to complete a timely diagnostic(Asnaoui et al., 2020; Chandra et al., 2021). To combat this situation some Computer-Aided Diagnosis (CAD) systems are the need of the time.

In Computer-Aided Diagnosis (CAD) using X-ray images, machine learning has already been applied successfully in many clinical and radiological studies to describe various characteristics of radio-imaging (Jaiswal et al., 2019; Pesce et al., 2019; Xue et al., 2018). Also to diagnose COVID-19 using X-ray images, a plethora of machine learning-based methods with varying sources and amount of training data have been proposed in the literature (Abbas et al., 2020; Ardakani et al., 2020; Asnaoui et al., 2020; Chowdhury et al., 2020; Jain et al., 2020; Minaee et al., 2020; Ozturk et al., 2020; Panwar et al., 2020; Toğaçar et al., 2020). Almost all the proposed methods in the literature are based on Convolution Neural Network (CNN) deep learning approach (LeCun et al., 1999). However, deep learning approaches to generalize well normally require an enormous amount of data which is not readily available right now. Although, some studies have been proposed using transfer learning where to fine-tune the model, a handsome amount of data is required keeping in view the dimensionality of features produced through pre-trained deep learning models. Moreover, the above studies have been proposed only to diagnose COVID-19 but its severity detection is still an open research question. To overcome these shortcomings, in this study we have proposed COVID-19 diagnosis and its severity predictor called COVIDX (COVId-19 Detection using X-ray images) by employing a blend of deep and shallow learning.

## 2. Methods

In this section, we discuss the details of our experimental strategy to design COVIDX (COVId-19 Detection using X-ray images) and its evaluation.

### 2.1. Datasets and their preprocessing

The datasets used in this study have been collected from different publicaly open online COVID-chest-X-ray image repositories (Cohen et al., 2020; Wang et al., 2020). We used two different datasets: i) dataset for COVID-19 diagnosis and ii) dataset for COVID-19 severity prediction. For COVID-19 diagnosis, we have a dataset of 576, 1583, and 4273 digital X-ray images (jpg format) of COVID-19 patients, healthy persons, and pneumonia patients, respectively. Whereas for COVID-19 severity prediction, we have a dataset of 114 and 164 X-ray images of highly severe and less severe COVID-19 patients, respectively. All the images in both datasets have been preprocessed including image resizing ($313 \times 313$ pixels), de-noising, and contrast stretching (Chandra and Verma, 2020b).

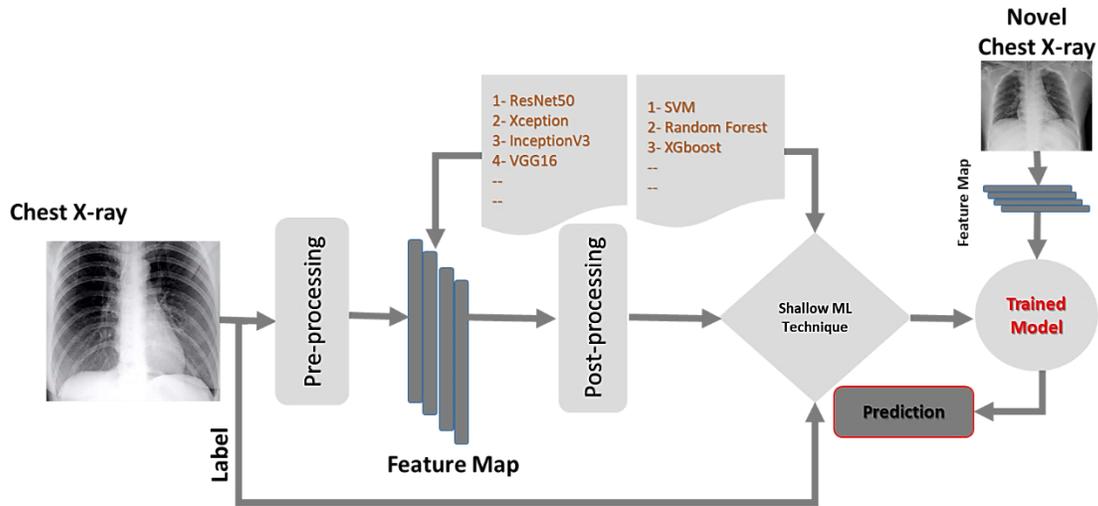

**Fig. 1.** A proposed methodology for the development of computer-aided COVID-19 diagnosis and its severity prediction system.

## 2.2. Proposed Methodology

We propose a machine learning approach for the identification and severity prediction of COVID-19 infection from raw digital chest X-ray images. The methodology adopted in this study has been depicted in Fig. 1.

## 2.3. Feature Extraction

To extract useful feature space from the digital chest X-ray images in our datasets, we have used different off-the-shelf CNN based pre-trained models on ImageNet. These pre-trained models include Resnet50 (He et al., 2015), InceptionV3 (Szegedy et al., 2015), Xception (Chollet, 2017), VGG16 (Simonyan and Zisserman, 2015), NASNetLarge (Zoph et al., 2018), DenseNet121 (Huang et al., 2018). The selection of these pre-trained CNN based models was based on their reported accuracy. Preprocessing and resizing required by the pre-trained models have been applied before extracting the features map.

### 2.4. Classifiers for the identification and prediction of severity of COVID-19

We propose a machine learning based approach for the identification and severity prediction of COVID-19 from digital chest X-ray images. As discussed earlier, the novelty of the proposed approach is that it uses a blend of pre-trained CNN based models to extract features and shallow learning algorithms for the classification purpose. The proposed scheme is based on the paradigm of transfer learning.

In this work, our dataset consists of examples of the form $(I_i, y_i)$ where $I_i$ is a chest X-ray image and $y_i \in \{+1, -1\}$ is its associated label. For COVID-19 identification, $y_i$ indicate whether $I_i$ COVID-19 (+1) or not (-1) and for COVID-19 Severity prediction $y_i$ indicate high (+1) or low (-1). For a given chest X-ray image $I_i$, we extract deep feature maps from it which are denoted by $x_i$. Our objective is to learn three separate functions for the identification and its severity prediction. For this purpose, we have used three different classifiers: classical Support Vector Machine (SVM), Random Forest (RF), and Gradient Boosting Machine (XGBoost) (Breiman, 2001; Cortes and Vapnik, 1995; Friedman, 2001).

#### 2.4.1. Support Vector Classification (SVC)

We used SVM for the diagnosis of COVID-19 and its severity prediction by learning a function $f(x) = \langle w, x \rangle$ with $w$ as parameters to be learned from the training data $\{(x_i, y_i) | i = 1, 2, \ldots, N\}$. The optimal value of the $w$ is obtained in SVM by solving the following optimization problem (Cortes and Vapnik, 1995).

$$min_{w,\xi} \frac{1}{2}\lambda\|w\|^2 + \sum_{i=1}^{N} \xi_i$$

$$\text{Such that for all } i: y_i\langle w, x_i \rangle \geq 1 - \xi_i, \xi_i \geq 0 \quad (1)$$

The objective function in Eq. (1) maximizes the margin while minimizing margin violations (or slacks ξ) (Cortes and Vapnik, 1995). The hyperparameter $\lambda = \frac{1}{C}$ controls the tradeoff between margin maximization and margin violation. We used both linear and radial basis function (RBF) kernels and coarsely optimized the values of λ and γ using grid search with scikit-learn (version:0.23) (Pedregosa et al., 2011).

### 2.4.2. Random Forest Classification (RFC)

Random Forest Classification (RFC) uses an ensemble learning technique based on bagging for regression. A random forest operates by constructing several decision trees in parallel during training and outputs the mean of the classes as the prediction of all trees(Breiman, 2001). It usually performs better on problems having features with non-linear relationships. Each classification tree in the RF is constructed on randomly sampled subsets of input features. In this study, we have optimized RF for the number of decision trees in the forest, the maximum number of features considered for splitting a node, the maximum number of levels in each decision tree, and a minimum number of samples required to split. We have also seen this regression technique effectively in action in many other studies (Abbasi et al., 2017; Ballester and Mitchell, 2010; Li et al., 2014; Moal et al., 2011).

### 2.4.3. XGBoost Calssification (XGBC)

XGBoost is also an ensemble learning technique based on the boosting that combines weak learners into a strong learner in an iterative fashion (Chen and Guestrin, 2016; Friedman, 2001). We have used trees as default base learners in XGBoost ensembles. In this study, we have optimized XGBoost in terms of the learning rate, maximum depth, the number of boosting iterations, booster, and subsample ratio with a grid search and using a python-based package xgboost 0.7 (Chen and Guestrin, 2016).

## 2.5. Model validation, selection, and performance assessment

We have divided the preprocessed images into two sub-sets: train-test set (80%), held-out validation set (20%), and reported performance metrics on both the sub-sets. For the train-test set, we have used 10-fold cross-validation (CV). In 10-fold CV, we have shuffled images in our datasets and then split them into 10 groups. Ten models have been trained and evaluated with each group given a chance to be held out as the test set (Abbasi and Minhas, 2016). For the held-out validation set, we trained the classification models using the whole train-test set and tested on the validation set. For performance metrics, we have used the area under the ROC curve (ROC), the area under the precision-recall curve (PR), and F-measure as performance measures for model evaluation and performance assessment(Abbasi and Minhas, 2016; Davis and Goadrich, 2006; Tharwat, 2020). We used grid search over training data to find the optimal values of hyper-parameters of different classification models.

## 2.6. Webserver for COVIDX

Table 1: Predictive performance for COVID-19 diagnosis across different classification models and feature maps using 10-fold CV (Healthy vs Un-healthy X-ray Images)

| Feature Map | SVC | | | RFC | | | XGBC | | |
|---|---|---|---|---|---|---|---|---|---|
| | ROC | PR | F1 | ROC | PR | F1 | ROC | PR | F1 |
| **Resnet50** | 0.98±0.01 | 0.98±0.01 | 0.98 | 0.99±0.03 | 0.98±0.08 | 0.96 | 0.98±0.03 | 0.98±0.08 | 0.98 |
| **Xception** | 0.99±0.01 | 0.98±0.01 | 0.99 | 0.99±0.01 | 0.99±0.01 | 0.97 | 0.99±0.01 | 0.99±0.01 | 0.98 |
| **InceptionV3** | 0.98±0.01 | 0.99±0.01 | 0.97 | 0.99±0.01 | 0.99±0.02 | 0.97 | 0.99±0.02 | 0.99±0.01 | 0.97 |
| **VGG16** | **0.99±0.01** | **0.99±0.01** | **0.99** | 0.98±0.01 | 0.97±0.01 | 0.98 | **0.99±0.01** | **0.99±0.01** | **0.98** |
| **NASNetLarge** | 0.98±0.01 | 0.98±0.01 | 0.96 | 0.99±0.01 | 0.99±0.01 | 0.97 | 0.98±0.03 | 0.98±0.01 | 0.96 |
| **DenseNet121** | 0.99±0.01 | 0.99±0.01 | 0.99 | **0.99±0.01** | **0.99±0.01** | **0.98** | 0.96±0.03 | 0.96±0.01 | 0.97 |

ROC (Area under the ROC curve), PR (Area under the precision-recall curve), F1 (F1 Score), SVC (Support Vector classifier), RF (Random Forest classifier), XGBC (XGBoost classifier). Bold-faced values indicate the best performance for each model.

We have developed and deployed a webserver of COVIDX which uses the optimal machine learning model for COVID-19 diagnosis and its severity prediction. This webserver takes a chest X-ray image and performs COVID-19 diagnosis and its severity prediction for it. After the successful submission of an X-ray image, the users are redirected to a page showing COVIDX predictions for COVID-19 diagnosis and its severity prediction. This process has broken into three steps: 1) whether the uploaded image belongs to a healthy person or unhealthy one (Healthy vs Un-healthy), 2) if the uploaded image belongs to an unhealthy person then whether it is COVID-19 infection or not (COVID-19 or not), and 3) if it is COVID-19 infection then how severe it is. The webserver is available at https://sites.google.com/view/wajidarshad/software.

## 3. Results and discussion

In this work, we have proposed a machine learning-based computer-aided COVID-19 diagnosis and its severity prediction. We have divided the COVID-19 diagnostic task into three sub-tasks based on the available data: 1) classification of healthy vs unhealthy, 2) classification of COVID-19 vs Pneumonia, and 3) COVID-19 severity prediction. For this purpose, we have used various machine learning algorithms and different deep features. In what follows we present results showing the prediction performance of our proposed method over cross-validation and an external test dataset.

**Table 2: Predictive performance for COVID-19 diagnosis across different classification models and feature maps on external validation dataset (Healthy vs Un-healthy X-ray Images)**

| Feature Map | SVC | | | RFC | | | XGBC | | |
|---|---|---|---|---|---|---|---|---|---|
| | ROC | PR | F1 | ROC | PR | F1 | ROC | PR | F1 |
| **Resnet50** | 0.97 | 0.97 | 0.91 | 0.98 | 0.97 | 0.91 | 0.97 | 0.96 | 0.93 |
| **Xception** | 0.98 | 0.97 | 0.91 | 0.97 | 0.97 | 0.91 | 0.97 | 0.97 | 0.93 |
| **InceptionV3** | 0.96 | 0.95 | 0.89 | 0.97 | 0.98 | 0.88 | 0.97 | 0.98 | 0.90 |
| **VGG16** | 0.97 | 0.97 | 0.91 | 0.98 | 0.97 | 0.91 | 0.98 | 0.97 | 0.91 |
| **NASNetLarge** | 0.96 | 0.95 | 0.87 | 0.96 | 0.96 | 0.85 | 0.96 | 0.96 | 0.88 |
| **DenseNet121** | **0.99** | **0.98** | **0.98** | **0.98** | **0.97** | **0.94** | **0.99** | **0.98** | **0.96** |

ROC (Area under the ROC curve), PR (Area under the precision-recall curve), F1 (F1 Score), SVC (Support Vector classifier), RF (Random Forest classifier), XGBC (XGBoost classifier). Bold-faced values indicate the best performance for each model.

### 3.1. Predictive performance for the classification of healthy versus unhealthy X-ray images

We have trained various shallow machine learning models for the classification of healthy versus unhealthy X-ray images with a range of deep learning-based feature maps and evaluated using both 10-fold cross-validation (CV) and on an external validation dataset. The results of our evaluation in both settings are shown in Tables 1 & 2 and Fig. 2 (A&B). Using 10-fold CV, we observed a maximum F1-score of 0.99 along with 0.99, and 0.99 as the area under the ROC curve, and the area under the PR curve, respectively with Support Vector Classifier and DenseNet121 feature map (Table 1). These results On 10-fold CV show the comparable performance of our proposed method in comparison to the state-of-the-art method proposed by Chandra et al (Chandra et al., 2021). Meanwhile, using an external dataset for the evaluation of our machine learning models trained on the training set, we observed a similar trend of performance with an F1-score of 0.98 along with 0.99, and 0.99 as the area under the ROC curve, and the area under the PR curve, respectively with Support Vector Classifier and DenseNet121 feature map (Table 2, Fig. 2A, Fig. 2B). These results show a better performance of our proposed method in comparison to the state-of-the-art method proposed by Chandra et al., with an F1-score of 0.96 and the area under the ROC curve of 0.93 with SVM (RBF kernel). The significantly better performance of SVM in comparison to Random Forest (RF) and Extreme Boosting Machine (XGB) is attributed to its ability to deal well with high dimensional features with fewer examples as in our case. Moreover, features extracted through pre-trained deep learning models perform better than handcrafted ones as in the case of the study performed by Chandra et al., (Chandra et al., 2021).

## 3.2. Predictive performance for the classification of COVID-19 versus Pneuminia X-ray images

We have also trained various shallow machine learning models for the classification of COVID-19 versus Pneumonia X-ray images with a range of deep learning-based feature maps and evaluated using both 10-fold cross-validation (CV) and on an external validation dataset. The results of our evaluation in both settings are shown in Tables 3 & 4 and Fig. 2 (C&D). Using 10-fold CV, we observed a maximum F1-score of 0.99 along with 0.99, and 0.99 as the area under the ROC curve, and the area under the PR curve, respectively with Support Vector Classifier and DenseNet121 feature map (Table 3). These results across 10-fold CV show an improved performance of our proposed method in comparison to the state-of-the-art method proposed by Chandra et al., with a maximum F1-score of 0.84 and the area under the ROC curve of 0.85 with SVM (linear kernel) (Chandra et al., 2021). Meanwhile, using an external dataset for the evaluation of our machine learning models trained on the training set we observed a similar trend of performance with an F1-score of 0.99 along with 0.99, and 0.99 as the area under the ROC curve,

**Table 3: Predictive performance for COVID-19 diagnosis across different classification models and feature maps using 10-fold CV (COVID-19 vs Pneumonia X-ray Images)**

| Feature Map | SVC | | | RFC | | | XGBC | | |
|---|---|---|---|---|---|---|---|---|---|
| | ROC | PR | F1 | ROC | PR | F1 | ROC | PR | F1 |
| **Resnet50** | 0.98±0.01 | 0.97±0.01 | 0.97 | 0.99±0.03 | 0.98±0.08 | 0.96 | 0.98±0.03 | 0.98±0.08 | 0.98 |
| **Xception** | 0.98±0.01 | 0.97±0.01 | 0.98 | 0.99±0.01 | 0.99±0.01 | 0.97 | 0.99±0.01 | 0.99±0.01 | 0.98 |
| **InceptionV3** | 0.98±0.01 | 0.97±0.01 | 0.97 | 0.99±0.01 | 0.99±0.02 | 0.97 | 0.99±0.02 | 0.99±0.01 | 0.97 |
| **VGG16** | 0.98±0.01 | 0.97±0.01 | 0.98 | 0.98±0.01 | 0.97±0.01 | 0.98 | 0.99±0.01 | 0.99±0.01 | 0.98 |
| **NASNetLarge** | 0.97±0.01 | 0.97±0.01 | 0.96 | 0.99±0.01 | 0.99±0.01 | 0.97 | 0.98±0.03 | 0.98±0.01 | 0.96 |
| **DenseNet121** | **0.99±0.01** | **0.99±0.01** | **0.99** | **0.99±0.01** | **0.99±0.01** | **0.98** | **0.99±0.03** | **0.99±0.01** | **0.99** |

ROC (Area under the ROC curve), PR (Area under the precision-recall curve), F1 (F1 Score), SVC (Support Vector classifier), RF (Random Forest classifier), XGBC (XGBoost classifier). Bold-faced values indicate the best performance for each model.

Table 4: Predictive performance for COVID-19 diagnosis across different classification models and feature maps on an external validation dataset (COVID-19 vs Pneumonia X-ray Images)

| Feature Map | SVC | | | RFC | | | XGBC | | |
|---|---|---|---|---|---|---|---|---|---|
| | ROC | PR | F1 | ROC | PR | F1 | ROC | PR | F1 |
| **Resnet50** | 0.99 | 0.99 | 0.98 | 0.99 | 0.98 | 0.96 | 0.99 | 0.98 | 0.98 |
| **Xception** | 0.99 | 0.99 | 0.98 | 0.97 | 0.97 | 0.95 | 0.97 | 0.97 | 0.95 |
| **InceptionV3** | 0.99 | 0.99 | 0.98 | 0.97 | 0.98 | 0.96 | 0.97 | 0.98 | 0.96 |
| **VGG16** | 0.99 | 0.99 | 0.99 | 0.98 | 0.97 | 0.91 | 0.98 | 0.97 | 0.94 |
| **NASNetLarge** | 0.98 | 0.99 | 0.98 | 0.96 | 0.96 | 0.90 | 0.96 | 0.96 | 0.93 |
| **DenseNet121** | **0.99** | **0.99** | **0.99** | **0.99** | **0.99** | **0.98** | **0.99** | **0.99** | **0.99** |

ROC (Area under the ROC curve), PR (Area under the precision-recall curve), F1 (F1 Score), SVC (Support Vector classifier), RF (Random Forest classifier), XGBC (XGBoost classifier). Bold-faced values indicate the best performance for each model.

and the area under the PR curve, respectively with Support Vector Classifier and DenseNet121 feature map (Table 4, Fig. 2C, Fig. 2D). These results show a better performance of our proposed method in comparison to the state-of-the-art method proposed by Chandra et al., with an F1-score of 0.91 and the area under the ROC curve of 0.91 even after using majority voting (Chandra et al., 2021). The significantly better performance of SVM in comparison to Random Forest (RF) and Extreme Boosting Machine (XGB) is attributed to its ability to deal well with high dimensional features with fewer examples as in our case. Moreover, features extracted through pre-trained deep learning models perform better than handcrafted ones as in the case of the study performed by Chandra et al., (Chandra et al., 2021).

### 3.3. Predictive performance for COVID-19 severity prediction

Along with COVID-19 identification, we have also tried to predict the severity of COVID-19 infection. For this purpose, we have trained several shallow machine learning models for the

Table 5: Predictive performance for COVID-19 severity prediction across different classification models and feature maps using 10-fold CV

| Feature Map | SVC | | | RFC | | | XGBC | | |
|---|---|---|---|---|---|---|---|---|---|
| | ROC | PR | F1 | ROC | PR | F1 | ROC | PR | F1 |
| **Resnet50** | 0.92±0.12 | 0.82±0.28 | 0.86 | 0.85±0.23 | 0.73±0.31 | 0.80 | **0.96±0.11** | **0.89±0.17** | **0.90** |
| **Xception** | 0.87±0.14 | 0.74±0.28 | 0.82 | 0.81±0.25 | 0.67±0.29 | 0.78 | 0.86±0.28 | 0.78±0.29 | 0.82 |
| **InceptionV3** | 0.86±0.13 | 0.68±0.29 | 0,74 | 0.73±0.30 | 0.58±0.30 | 0.75 | 0.82±0.26 | 0.76±0.29 | 0.78 |
| **VGG16** | 0.90±0.24 | 0.88±0.26 | 0.83 | 0.81±0.24 | 0.70±0.24 | 0.77 | 0.82±0.26 | 0.72±0.31 | 0.80 |
| **NASNetLarge** | 0.80±0.31 | 0.76±0.30 | 0.76 | 0.80±0.26 | 0.65±0.30 | 0.76 | 0.79±0.26 | 0.68±0.31 | 0.75 |
| **DenseNet121** | **0.96±0.17** | **0.90±0.19** | **0.90** | **0.90±0.19** | **0.78±0.26** | **0.84** | 0.90±0.20 | 0.83±0.26 | 0.90 |

ROC (Area under the ROC curve), PR (Area under the precision-recall curve), F1 (F1 Score), SVC (Support Vector classifier), RF (Random Forest classifier), XGBC (XGBoost classifier). Bold-faced values indicate the best performance for each model.

classification of more severe versus less severe COVID-19 infection using X-ray images of COVID-19 infected patients with a range of deep learning-based feature maps and evaluated using 10-fold cross-validation (CV). To the best of our knowledge, this is the first study on COVID-19 Severity prediction. The results of our evaluation are shown in Table 5. Using 10-fold CV, we observed a maximum F1-score of 0.90 along with 0.96, and 0.90 as the area under the ROC curve, and the area under the PR curve, respectively with Support Vector Classifier and DenseNet121 feature map (Table 3). These results show a reasonable performance of COVIDX over the severity prediction of COVID-19 infection.

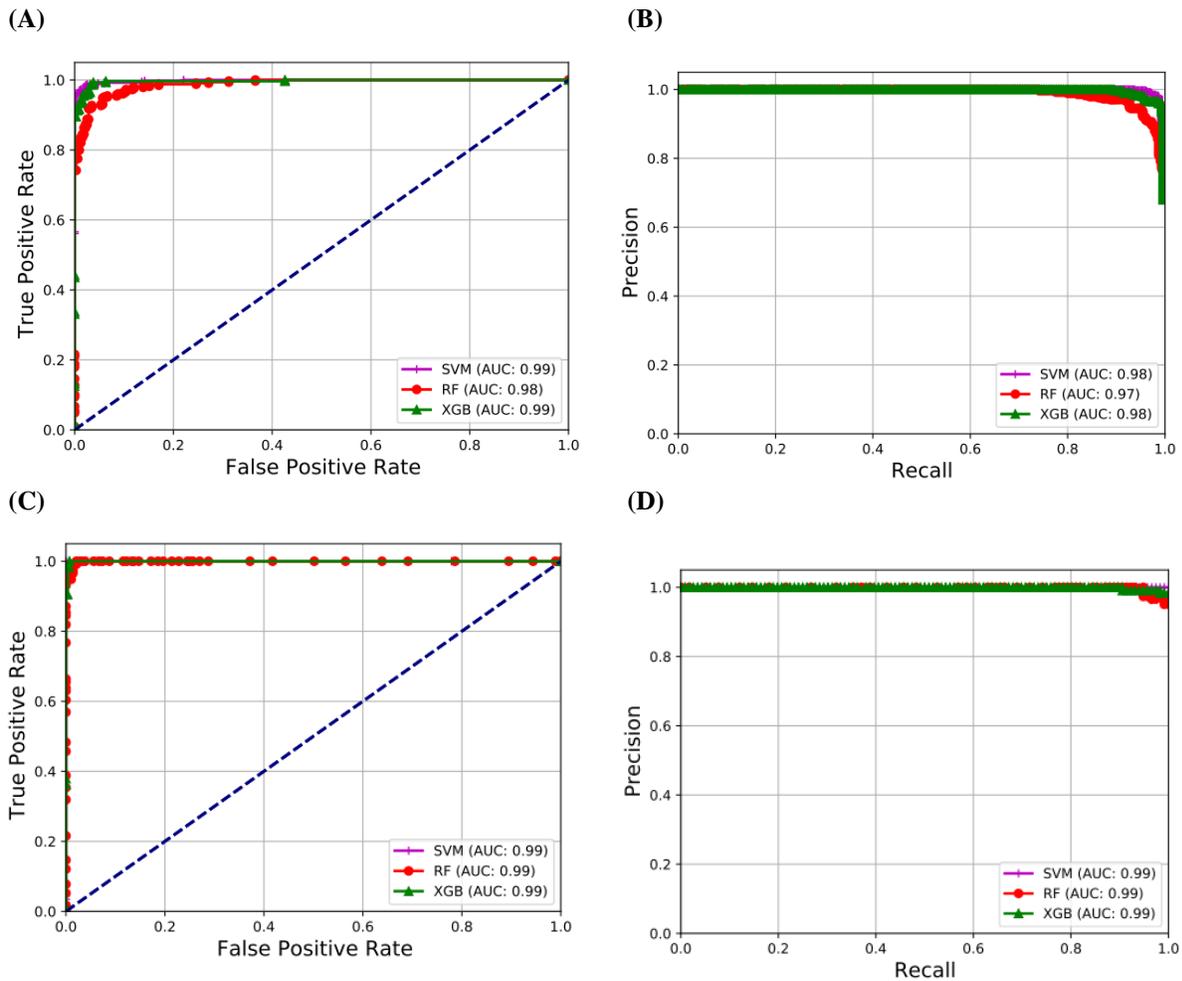

**Fig. 2.** Receiver Operating Characteristic (ROC) and Precision-Recall (PR) curves showing predictive performance of COVIDX for the classification of chest X-ray images (Healthy vs Unhealthy and COVID-19 vs Pneumonia) across different classifiers (SVM, RF, XGB) and DenseNet feature map on an external validation dataset. **Healthy vs Unhealthy** (A&B); **COVID-19 vs Pneumonia** (C&D).

## 3.4. Performance of proposed computer-aided COVID-19 diagnostic system in a real setting

Performance of COVIDX has also been evaluated using its webserver in a real setting under the supervision of an experienced radiologist at Abbas Institute of Medical Sciences (AIMS) hospital located in Muzaffarabad, Azad Jammu & Kashmir, Pakistan. For this purpose, 30 X-ray images (anonymous) belonging to different classes (10 COVID-19 infected, 10 Pneumonia infected, and 10 Healthy persons) have been used. Results obtained through this evaluation are shown as a confusion matrix in Fig. 3. Our proposed system (COVIDX) has been able to classify correctly 9 out of 10 provided X-ray images of COVID-19 infected patients and 1 as Pneumonia (Fig. 3). Similarly, for the provided X-ray images of Pneumonia infected patients, our system classified correctly 8 out of 10 images, 1 as COVID-10 and 1 as healthy (Fig. 3). For provided health x-ray images, our system classified 9 out of 10 images correctly as healthy and 1 as Pneumonia. These results justify the use of our proposed system in real settings.

## 4. Conclusion

In the present study, we have designed and developed a system called COVIDX for the preliminary diagnosis of COVID-19 infection and its severity prediction from a raw X-ray image. The performance evaluation through 10-fold CV, on an external validation dataset, and in a real setting show that our proposed system can efficiently be used to diagnose COVID-19 infected patients and to suggest precautionary measures (such as quarantine and RT-PCR test) avoiding the further surge of the infection. The key findings of the study are listed as follows:

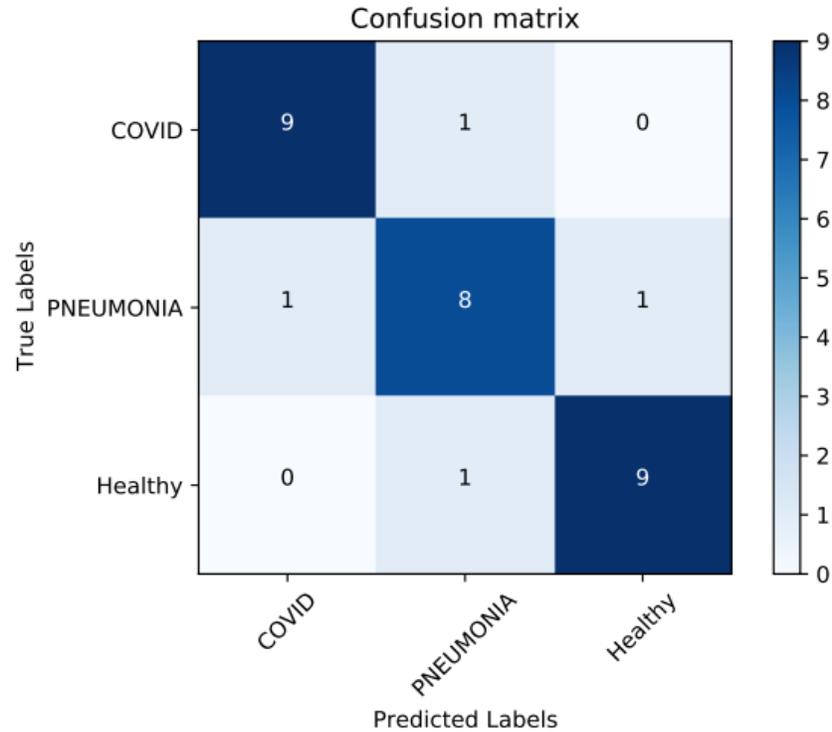

**Fig. 3. Confusion matrix**: Showing COVIDX performance in a real setting used by an experienced radiologist.

- Our proposed system performed significantly better in comparison to the state-of-the-art existing systems during the rigorous adopted evaluation criteria even in the presence of substantial variations in the input CXR images.

- Our proposed system not only diagnose COVID-19 but also predict its severity.

- We have made our proposed system accessible through a publically open cloud-based webserver and open-source code.

**Acknowledgments**

The authors would like to acknowledge the services of all those who have provided compiled and annotated X-ray image dataset as open access repositories.

**Availability of data and materials**

All data generated or analyzed during this study are included in this paper or available at online repositories. A Python implementation of the proposed method together with a webserver is available at https://sites.google.com/view/wajidarshad/software and https://github.com/wajidarshad/covidx.

**Authors' contributions**

WAA conceived the idea, developed the scientific workflow, performed the experiments, analyzed and interpreted the results, and was a major contributor in manuscript writing. SAA contributed to the analysis of the results and writing of the manuscript. SA helped in results interpretation and validation, formal analysis, and manuscript writing. All authors have read and approved the final manuscript.

**Competing interests**

The authors declare that they have no competing interests.